\documentclass{vgtc}                          

\ifpdf
  \pdfoutput=1\relax                   
  \usepackage{graphicx}                
  \DeclareGraphicsExtensions{.pdf,.png,.jpg,.jpeg} 
\else
  \usepackage{graphicx}                
  \DeclareGraphicsExtensions{.eps}     
\fi%

\graphicspath{{figures/}{pictures/}{images/}{./}} 

\usepackage{microtype}                 
\PassOptionsToPackage{warn}{textcomp}  
\usepackage{textcomp}                  
\usepackage{mathptmx}                  
\usepackage{times}                     
\usepackage{cite}                      
\usepackage{tabu}                      
\usepackage{booktabs}                  
\usepackage[usenames, dvipsnames]{color}

\usepackage{mathtools}

\DeclarePairedDelimiter{\floor}{\lfloor}{\rfloor}

\usepackage{verbatim}
\usepackage{url}
\usepackage{enumitem}
\usepackage{wrapfig}
\usepackage{amsmath}
\usepackage{xcolor}
\usepackage{stfloats}
\onlineid{0}

\vgtccategory{Research}

\vgtcinsertpkg

\title{Virtual Reality Lensing for Surface Approximation in Feature-driven DVR}

\author{Roberta Mota\thanks{e-mail: roberta.cabralmota@ucalgary.ca} %
\and Ehud Sharlin\thanks{email: ehud.sharlin@ucalgary.ca} %
\and Usman Alim\thanks{email: u.alim@ucalgary.ca}}
\affiliation{\scriptsize Department of Computer Science, University of Calgary, Canada}

\abstract{
\vspace{-3px}
We present a novel lens technique to support the identification of heterogeneous features in direct volume rendering (DVR) visualizations.
In contrast to data-centric transfer function (TF) design, our image-driven approach enables users to specify target features directly within the visualization using deformable quadric surfaces. 
The lens leverages quadrics for their expressive yet simple parametrization, enabling users to sculpt feature approximations by composing multiple quadric lenses.
By doing so, the lens offers greater versatility than traditional rigid-shape lenses for selecting and bringing into focus features with irregular geometry.
We discuss the lens visualization and interaction design, advocating for bimanual spatial virtual reality (VR) input for reducing cognitive and physical strain.
We also report findings from a pilot qualitative evaluation with a domain specialist using a public asteroid impact dataset.
These insights not only shed light on the benefits and pitfalls of using deformable lenses but also suggest directions for future research.
} 

\CCScatlist{
  \CCScatTwelve{Human-centered computing}{Human computer interaction}{Interaction paradigms}{Virtual Reality};
  \CCScatTwelve{Human-centered computing}{Visu\-al\-iza\-tion}{Visualization Application Domains}{Scientific Visualization}
}

\begin{document}


\maketitle

\vspace{-3px}
\section{Introduction}
\vspace{-3px}

As volume data is prevalent across numerous scientific fields---\textit{e.g.}, engineering, medical imaging, and astrophysics, DVR techniques have been proposed to portray the whole 3D data within a single 2D image \cite{Hansen:2011}. A widely-used DVR algorithm is ray casting, in which data values are accumulated along pixel-wise rays and later mapped into visual encodings composing the final 2D projection image \cite{Roettger:2003}.
However, since tens to hundreds of data values are compressed along each view ray into a single pixel, volume rendering often suffers from occlusion and visual clutter \cite{Hurter:2014}.
Consequently, identifying features of interest located inside the volume can become challenging. This is particularly true for \textit{heterogeneous features}---features composed of varying data value ranges, which are difficult to distinguish from surrounding context sharing similar intensity values.

To mitigate clutter and occlusion, focus+context techniques for volume feature enhancement have been proposed---wherein volume data inside and outside the focus area are rendered using distinct visual styles.
These approaches can be classified as either \textit{data-driven} or \textit{image-centric} \cite{Pfister:2001}.
The former commonly uses scalar fields and derived properties when designing TFs to put focus on volume features:
users first identify data ranges corresponding to voxels belonging to the feature, and then assign visual TF parameters (\textit{e.g.} color, opacity) to these intensity ranges. 
Despite their widespread use, designing TFs is often unintuitive due to two main issues: i) inaccurate identification and isolation of intensity ranges, and ii) indirect relation between the TF parameters and the resulting visualization.
TF design therefore often requires repetitive manual adjustments to find optimal TF parameters that clearly depict the target feature---and this becomes even more cumbersome when dealing with heterogeneous features \cite{Pfister:2001}.
However, these issues may be alleviated through image-centric approaches; rather than tuning indirect TF parameters, users specify features in focus by directly interacting with the visualization. 

\vspace{5px}
\noindent \textbf{Contribution} Given the aforementioned, this paper contributes an \textbf{immersive lens technique} for the identification of heterogeneous features in DVR visualizations. 
The lens incorporates a deformable surface geometry to closely match a portion of the feature of interest, based on the premise that a surrogate of the feature can be constructed as a composition of multiple simple surfaces.
Furthermore, due to the inherent spatial nature of volumetric data, we employed VR technology to provide an associative 3D interaction space and bimanual input, with the ultimate goal of reducing physical and cognitive strain through more intuitive interaction mappings.
By doing so, our lens seeks to address the two key limitations of TF design. First, users no longer need to precisely isolate intensity values; instead, they sculpt a rough approximate representation of the target feature. Second, TF parameter tuning is unnecessary, as visual focus is directed toward the user-defined proxy geometry.

\vspace{3px}
\noindent \textbf{Outline} This paper is structured as follows: Section \ref{sec:relatedWork} reviews related work on focus+context techniques in DVR and general 3D visualizations. Section \ref{sec:lensConcept} presents the lens concept, while Section \ref{sec:lensTechnique} details its interaction and visual design. Section \ref{sec:expertFeedback} reports findings from a pilot qualitative evaluation and discusses opportunities for future research. Finally, Section \ref{sec:conclusion} concludes the paper.

\begin{figure*}[!t]
\centering
\includegraphics[width=1\linewidth]{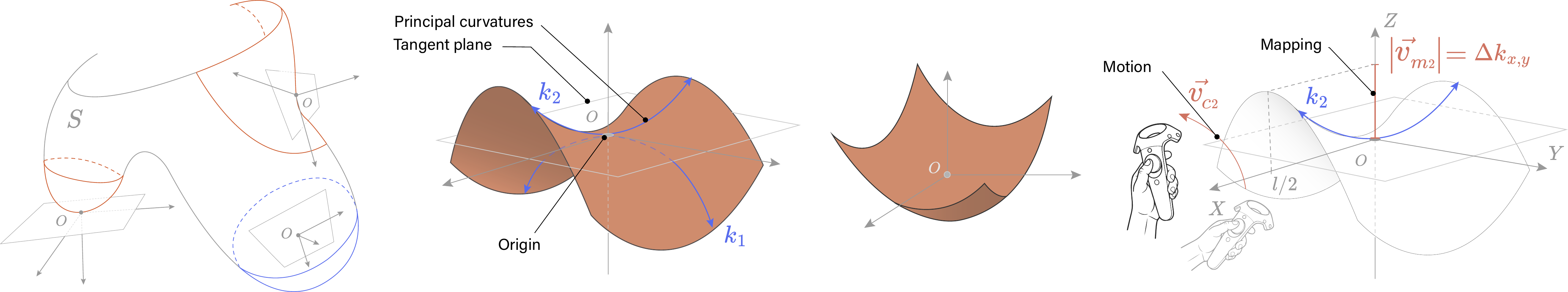}
\vspace{-0.57cm}
\caption{Our premise is that each point on $S$ has a neighborhood that can be represented as $z = h(x, y)$, where $h$ is a differentiable function (\textit{a}) \cite{doCarmo:2005}. Quadric variations: hyperbolic paraboloid ($k_{1}k_{2} < 0$) (\textit{b}), and ellipitic paraboloid ($k_{1}k_{2} > 0$ and $k_{1} \neq k_{2}$) (\textit{c}).Constrained motion mapping along the $z$-axis during curvature manipulation (\textit{d}).}
\vspace{-0.5cm}
\label{fig:lensConcept}
\end{figure*}

\vspace{-3px}
\section{Related Work}
\label{sec:relatedWork}
\vspace{-4px}

This section examines existing focus+context techniques aimed at supporting feature identification in DVR and more general 3D visualizations.
For a more in-depth review of lens-based techniques, we refer the reader to previous survey literature \cite{mota2025, tominski2017}.

\vspace{-3px}
\subsection{Focus+Context in Volume Rendering}
\vspace{-3px}

Focus+context techniques to assist users locate target features in DVR visualizations can be roughly segmented into two groups: data-centric transfer functions and image-centric techniques.

\vspace{5px}
\noindent \textbf{Data-centric transfer functions} The most common transfer functions are \textit{1D}, mapping scalar data values to visual encodings---\textit{e.g.} color or transparency. They are widely used in popular visualization software like Paraview \cite{Paraview}.
However, elementary 1D TFs perform poorly when a target feature cannot be easily separated (isolated) from their surrounding, occluding context with similar intensity values \cite{Preim:2016}.
\textit{Higher-order} transfer functions incorporating gradient and curvature information have been proposed to better isolate features of interest in volumetric data \cite{Kniss:2002,Kindlmann:2003, Kindlmann:1998, Levoy:1988}.
Like 1D TFs, these higher-order functions commonly employ histograms to display  distributions across attribute dimensions. However, as dimensionality increases, the resulting histograms become increasingly challenging to interpret and manipulate.
Despite their significance to the quality of focus+context volumetric images, designing good transfer functions still proves difficult. The indirect relation between TF parameters and visual outcomes makes their design unintuitive, often requiring time-consuming trial-and-error adjustments until meeting a quality image---this is further exacerbated as small TF parameter changes lead to disproportionate alterations in the visualization \cite{Mindek:2017}.

\vspace{3px}
\noindent
\textbf{Image-centric techniques} Some techniques are image-driven, allowing users to directly manipulate visual elements without being exposed to the underlying TF parameters.
There are \textit{semi-automated} methods using numerical or topological information to guide TF specification, so that spatially-connected features are jointly distinguished. However, they require data presegmentation, which is computationally costly and offers limited user interactivity to control what gets selected and what does not  \cite{Guo:2013, Marks:1997}. Li \textit{et al}. tackled virtual luggage unpacking via a confidence-driven volume segmentation that recursively splits connected regions until semantically meaningful objects are obtained \cite{Li:2012}.
Other techniques offer more \textit{interactive} controls, relying on user input to guide TF design \cite{Guo:2011, Luo:2018}.
Sompagnimdi and Hurter proposed two brushing methods for 3D baggage scans: range-based and magic brushing, which selectively removes DVR voxels based on user-defined density thresholds  \cite{Traoré:2017}.
While these methods enhance user control, they struggle when target and occluder structures share similar scalar values. They may also be constrained to 2D screen-space interactions, reducing effectiveness in intricate datasets.
Closer to our work, Traoré \textit{et al.} addressed similar-density occluders and targets by proposing a lens that maintains the volume context while revealing partially-hidden targets \cite{Traoré:2018}. The lens first pushes occludeing voxels radially outward, then applies a fish-eye deformation to expand the focus area.
However, the technique assumes users can reliably identify and select partially-visible targets. It may also fail to separate features that are in close proximity to (\textit{i.e.} cannot easily be spatially isolated from) occluders.

\vspace{-3px}
\subsection{Lenses in 3D Visualization}
\vspace{-3px}

Lenses are a specific class of focus+context techniques, defined as a spatial selection through which a base visualization is modified to provide an alternative visual representation of the data in focus \cite{mota2025}.
In addition to position, orientation, and scale, the \textit{shape} of the lens is a key parameter influencing the spatial selection of the data brought into focus.
Lens shapes can be classified according to their dimensionality as 2D, 2.5D, or 3D.

\vspace{5px}
\noindent \textbf{2D} Polygonal shapes are the most commonly used---\textit{e.g.} circular or square, which are placed and manipulated in screen space. Although 2D lenses have been used in 3D visualizations \cite{kister2015, kim2012embodied, chang2013, hurter2011}, their 2D nature constrains the lens manipulations to the view plane and leads to two major limitations when handling 3D data: an inability to carry out a 3D selection, and an absence of spatial correlation between the 2D lens alignment and the underlying 3D visualization.

\vspace{3px}
\noindent \textbf{2.5D} Some lenses  embedd 2D lenses in 3D spaces, thereby augmenting lens manipulations to 3D \cite{bichlmeier2009, nam2019, gasteiger2011}. While this enhances spatial correlation, it increases interaction effort to position and align the lenses according to the underlying data. In addition, the flat nature of 2.5D lenses constraint the selection to only a slice of the 3D data.

\vspace{3px}
\noindent \textbf{3D} There are lens shapes consisting of 3D volumes---\textit{e.g.} a sphere, which enable volumetric spatial selections of the 3D data \cite{trapp2008, johnson2019, borst2007volume}. Due to their 3D nature, the lens manipulations are prone to suffer from interaction effort similar to that of 2.5D lenses. Additionally, such lenses may not closely align with surface geometries due to their typically rigid shapes; this can lead to perceptual issues, such as the lens itself occluding parts of visualization—especially areas with intricate geometry.
In contrast, our approach follows in the footsteps of works like those by Steimle \textit{et al.} \cite{steimle2013flexpad} and Looser \textit{et al.} \cite{looser20073d}, exploring flexible, bendable surfaces to achieve greater expressiveness and versatility in selecting the feature in focus.

\vspace{5px}
\noindent Building on the aforementioned discussion, we investigate an image-centric lens technique designed to support the identification of heterogeneous features in DVR visualizations.
To this end, the lens shape not only matches the volumetric nature of the underlying data but also incorporates a deformable design. offering greater versatility and precision in selecting irregular features---and, ultimately, more accurately reflecting user intent than standard rigid 3D shapes.


\vspace{-3px}
\section{Lens Concept \& Definition}
\label{sec:lensConcept}
\vspace{-3px}

This paper introduces an interactive lens that brings into focus heterogeneous features in DVR visualizations.
To this end, our lens is conceptualized as a deformable 2D manifold that emulates a local surface patch,
based on the premise that a meaningful approximation of the target feature can be built by composing multiple simple patches---specifically, second-order parametric surfaces. As follows, we describe core concepts and definition of our quadric lens.

\vspace{5px}
\noindent \textbf{Definition} do Carmo has proven that any smooth surface can be \textit{locally} approximated by a differentiable function \cite{doCarmo:2005}. Given a point $p$ on a surface $S$, we can define an orthogonal Cartesian coordinate system $xyz$ such that the $xy$-plane coincides with the tangent plane of $S$ at $p$, and the $z$-axis aligns with the normal at $p$ (\textcolor{black}{Fig. \ref{fig:lensConcept}--\textit{a}}). It follows that, in the neighborhood of $p$, the surface $S$ can be represented in the form: 

\vspace{-12px}
\begin{align*}
\textbf{$r(x, y) = [x,\ y,\ h(x, \ y)]^T$}
\end{align*}
\vspace{-12px}

\noindent where $h$ is a differentiable function with $h(0, 0) = h_{x}(0,0) = h_{y}(0,0) = 0$. Assuming $p$ is non-planar, the surface $S$ can be locally approximated by a second-order parametric function:

\vspace{-8px}
\begin{align*}
\textbf{$r(x, y) = [x,\ y,\ \frac{1}{2} \times (k_{1} \times x^2 + \ k_{2} \times y^2)]^T$}
\end{align*}
\vspace{-8px}

\noindent The aforementioned equation defines a \textit{quadric surface} $Q$, where $k_{1}$ and $k_{2}$ parameters are the principal curvatures of the quadric---\textit{i.e.}, they measure the surface bending magnitude at each point $p \in Q$.
In this way, a variety of quadric surfaces can be modeled based on simple combinations of $k_{1}$ and $k_{2}$, as shown in \textcolor{black}{Fig. \ref{fig:lensConcept}--\textit{b, c}}.

\vspace{2px}
\noindent Due to their favorable compromise between simplicity and expressiveness, we investigated the use of quadric surfaces for our shape-adaptive lens design.
To afford deformable geometry, the lens quadric includes five interactive \textit{control points}: one at the origin and four at the extremities of the principal curvature directions---see \textcolor{black}{Fig. \ref{fig:lensConcept}--\textit{b}}. These curvature endpoints act as manipulable handles to dynamically deform the quadric surface, and are defined as follows:

\vspace{-4px}
\begin{align*}
\textbf{$c_{i,\ k_{1}} = \frac{1}{2} ((-1)^i \times l, \ 0, \ k_{1} \times (\frac{l}{2})^2)$}
\end{align*}
\vspace{-0.3in}

\begin{align*}
\textbf{$c_{i,\ k_{2}} = \frac{1}{2} (0, \ (-1)^i \times l, \ k_{2} \times (\frac{l}{2})^2)$}
\end{align*}
\vspace{-4px}

 \noindent where $l$ is the length of the quadric and $\ i \in \{1, 2\}$. Each mirrored pair of control points---reflected across the $x$- or $y$-axis---shares the same curvature value due to the quadric's axial symmetry.


\vspace{-3px}
\section{Lens Technique}
\label{sec:lensTechnique}
\vspace{-3px}

Grounded in our theoretical understanding and design of deformable quadric surfaces, in this section we report the rationale for our lens interaction and visualization design.

\vspace{-2px}
\subsection{Interaction Design}
\vspace{-2px}
The potential for direct, bimanual spatial manipulation led us to incorporate our lens technique in VR using an HTC Vive. Its room-scale tracking allows users to walk around or even inside the data, and seamlessly interact with it via a spatially-tracked headset and two controllers. In the following, we describe the interactions supported by our quadric lens---each denoted by \textbf{I$_{i}$}. In addition to standard spatial transformations, users can also interactively modify the quadric curvatures, thereby sculpting different variants of quadric surfaces.

\begin{figure}[t!]
\centering
\includegraphics[width=3.3in]{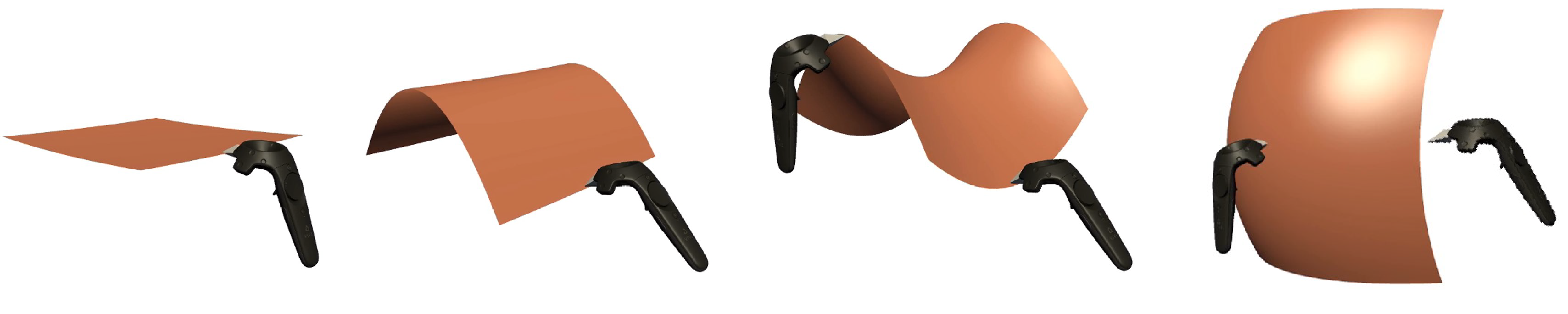}
\vspace{-0.3cm}
\caption{Curvature modulation begins by grabbing a control point (\textit{left}). While holding the point, z-constrained movements from the controller is mapped to the quadric curvature value (\textit{middle}). Scaling is performed by grabbing the lens with a controller and pressing-and-pulling with the other (\textit{right}).}
\label{fig:lensInteractions}
\vspace{-0.5cm}
\end{figure}

\begin{figure*}[!b]
\centering
\vspace{-0.2cm}
\includegraphics[width=1\linewidth]{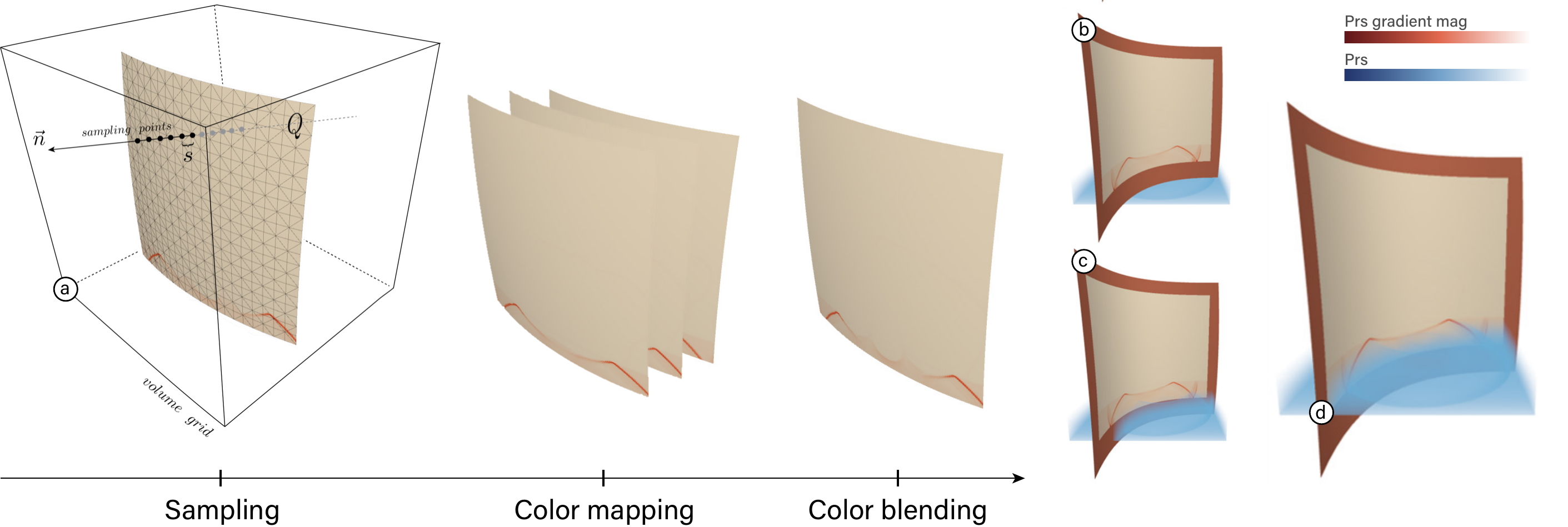}
\vspace{-0.6cm}
\caption{Illustrative sampling and color mapping and blending along the normals of a quadric surface $Q$ (\textit{a}). The three visualization designs: DVR with $z$-culling (\textit{b}), DVR with neighbouring $z$-cullling (\textit{c}), and standard DVR with additional alpha factor (\textit{d}).}
\label{fig:lensVisualizations}
\end{figure*}

\vspace{5px}
\noindent \textbf{I$_{1}$} \textit{Instantiation}. Users instantiate a lens by pressing the grip button on either controller.
\textbf{I$_{2}$} \textit{Positioning}. Similar to a mouse drag-and-drop interaction, we use a grab-and-release metaphor to position and orient the lens. The \textit{grabbing} metaphor enables direct manipulation in 3D spaces, and is known to be very easily understood due to its natural kinaesthetic correspondence---an isomorphism between the user's hand movement and the visually perceived motion \cite{Poupyrev:2001, Ware:1988}. By pressing-and-holding the trigger button near the origin control point on either controller, the user \textit{grabs} the lens. While held, the lens moves in one-to-one correspondence with the controller’s position and orientation, appearing to be rigidly attached to it. 
\textbf{I$_{3}$} \textit{Scaling}. While grabbing a quadric lens with a controller, users can scale it as a secondary action by pressing the trigger button on the other controller and pulling outward, as depicted in \textcolor{black}{Fig. \ref{fig:lensInteractions}--\textit{right}}. The scale factor is defined by the distance between the initial and current positions of the secondary controller. This design grants users the ability to seamlessly combine spatial manipulations without compromising interaction fluidity---grabbing the lens with one hand and scaling it by grabbing and pulling with the other. 
\textbf{I$_{4}$} \textit{Curvature modulation}. Users can also manipulate the quadric curvatures---either individually or jointly. When the trigger button is pressed on either controller, the press position determines the nearest curvature control point. While grabbing that point, its $z$-position is bound to the upward movement of the controller---see \textcolor{black}{Fig. \ref{fig:lensConcept}--\textit{d}}. 
Our design serves two purposes. First, to improve precision in spatial manipulation through a virtual constraint that ignores extraneous degrees of freedom in the user input \cite{mine1997isaac}. Second, to bring a sense of direct manipulation, as if the end-effector point were physically snapped to the controller, as shown in \textcolor{black}{Fig. \ref{fig:lensInteractions}--\textit{left, middle}}.
This comes from our observation that changes in $k_{1}$ and $k_{2}$ coefficients directly affect surface elevations along the $z$-axis. To maintain a consistent mapping between interaction and data spaces, we constrain the controller input to the upward axis---ignoring extraneous degrees of freedom. As a result, a grabbed control point translates along a constrained motion ($\Delta z = \Delta k$).

\vspace{-3px}
\subsection{Visual Design \& Implementation}
\vspace{-2px}

As follows, our lens visual design brings a feature of interest into \textit{focus}, while the underlying DVR provides \textit{context}.

\vspace{5px}
\noindent \textbf{Focus} The focus effect is rendered on the quadric mesh. For each vertex $v = (x, \ y, \ z)$ of the quadric surface, we sample the corresponding voxel $i_{v} = \floor[\big]{v \cdot D}$ within the volume grid, where $ D = (D_{x}, \ D_{y}, \ D_{z})$ represents the grid dimensions.
In addition, neighboring voxels are sampled along the surface normal direction, using a step size $s = \sqrt{3} / D$, as illustrated in Fig. \textcolor{black}{\ref{fig:lensVisualizations}--\textit{a}}. The sampled voxel values are then mapped to color using a colormap, and later blended together. Lastly, we apply ambient and diffuse lighting to enhance depth perception.
This local refinement through normal-based color blending is consistent with the quadric geometry and produces smoother color (\textit{i.e.} intensity) transitions as users manipulate the surface within the volume. This can facilitate feature detection by capturing sharp transitions between sequential voxels and allowing greater tolerance for imprecise surface placement relative to the feature.
Results for different sampling resolutions, rendered on a quadric surface using a diverging cool-to-warm colormap, are provided in the supplementary material. We observed that high number of samples may introduce visual artifacts. Determining the optimal sampling density $n$ remains an open question; in this work, we found $n = 5$ to yield satisfactory results.
Additionally, users can specify the attribute rendered on the quadric image, toggling between the scalar value and its gradient magnitude. We hypothesize users may initially rely on the gradient magnitude to guide feature discovery, then switch to scalar values to analyze intensities along the feature's focus surface.

\vspace{4px}
\noindent \textbf{Context} The DVR should provide surrounding context to the focus surfaces. To this end, a central design question was how to properly combine \textit{focus} and \textit{context}, emphasizing quadric images while minimizing visual clutter.
As shown in \textcolor{black}{Fig. \ref{fig:lensVisualizations}--\textit{b-d}}, we explored and developed three visualization techniques, each indexed \textbf{Vis\(_i\)}. 
\textbf{Vis\(_1\)} \textit{Standard DVR}. A standard direct volume rendering using an alpha parameter to adjust global transparency. 
\textbf{Vis\(_2\)} \textit{DVR with depth culling}. DVR samples are culled based on a buffer that stores the depths of all quadric surfaces. Sample points located in front of any surface are discarded. 
\textbf{Vis\(_3\)} \textit{DVR with neighboring depth culling}. Similar to the previous approach, but sampled points are culled only if they fall within a small depth neighborhood of the surfaces---in our case, $\Delta z \leq 0.01$. 
While all methods convey sufficient context, we favor neighboring culling since we believe it enhances both depth perception and the legibility of the focus visualization. 

\vspace{4px}
\noindent \textbf{Apparatus} The lens visualizations were implemented using Unity 3D \cite{Unity} along with an HTC Vive \cite{HTCVive}, which features a 2160×1200 display (1080×1200 per eye) at 90 Hz. Using an Intel® Core™ i3 CPU and a GeForce GTX Titan 6GB GPU, we achieve interactive frame rates of 45 FPS.

\begin{figure*}[!t]
\centering
\includegraphics[width=1\linewidth]{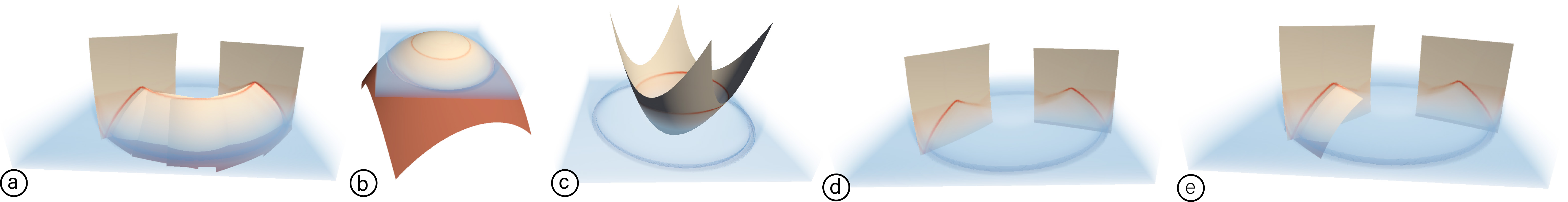}
\vspace{-0.5cm}
\caption{Feature surrogate examples from the expert feedback session (\textit{a--c}). Reference guides display boundary lines for the pressure shock wave (\textit{d}). The guides assist the placement of quadric surfaces (\textit{e}) that constitute an approximation of the feature (\textit{a})}
\vspace{-0.5cm}
\label{fig:lensEvaluation}
\end{figure*}


\vspace{-1px}
\section{Expert Feedback}
\label{sec:expertFeedback}
\vspace{-2px}

To demonstrate and evaluate the applicability of our lens, we conducted a pilot qualitative evaluation with a domain expert, a postdoctoral fellow with over 7 years of academic experience.
The researcher first described a case analysis, followed by a brief introduction on how to interact with the visualization. Throughout the study session, the expert was encouraged to ask questions and think aloud to express reasoning.
In this section, we begin by briefly describing the case analysis used in the feedback session, followed by the collected findings labeled as \textbf{F$_{i}$}.

\vspace{5px}
\noindent \textbf{Case analysis} The assessment was based on a public dataset of asteroid impact in deep ocean waters \cite{SciVisContest}.
Such events transfer enormous kinetic energy from the asteroid to the surrounding water, generating pressure waves that propagate radially from the impact site.
Analyzing how a shock wave travels is crucial for understanding its consequences---\textit{e.g.} tsunamis.
However, visualizing such feature in DVR is challenging. As the pressure pulse attenuates over distance, variations within the shock wave increasingly resemble the ambient ocean pressure, making it  difficult to isolate the heterogeneous shock wave from its surroundings (ocean) using TF design.
In this context, the domain expert was asked to isolate heterogeneous shock waves at different stages of propagation.

\vspace{6px}
\noindent \textbf{F$_{1}$} \textit{Multiple small $>$ single large for approximation and context.} A key takeaway from our feedback sessions was that better shape approximations were achieved by composing multiple, small quadric lenses rather than relying on a single, large one. Although a single quadric image could capture the entire feature, part of its surface was underutilized and introduced unnecessary visual clutter and occlusion on the surrounding volume---ultimately, reducing contextual awareness (\textcolor{black}{Fig. \ref{fig:lensEvaluation}--\textit{a-c}}).
\textbf{F$_{2}$} \textit{Role as alignment aids}.
Another insight from our feedback session was that users employed quadric lenses not only as proxy geometries but also as spatial guidelines. In these cases, quadric images were not used to approximate surface patches of the feature directly, but rather to assist in the placement and alignment of other quadric lenses---functioning similarly to ruler guides from Adobe Photoshop \cite{Photoshop}. 
These reference guides were often cross-sectional surfaces, strategically placed at locations where key structural lines of the feature were visible. A common workflow involved placing one or more of these guides first, then constructing the proxy shape by adjusting additional quadric lenses relative to them---see \textcolor{black}{Fig. \ref{fig:lensEvaluation}--\textit{d, e}}. 
\textbf{F$_{3}$} \textit{Locking against accidental edits}.
Another suggestion was the addition of a lock/release mechanism to prevent users from unintentionally selecting or editing quadric images that were already correctly positioned. We observed that the modeling process was sometimes disrupted by accidental interactions.
\textbf{F$_{4}$} \textit{Visual aid for control point disambiguation}.
It was also suggested providing visual feedback for candidate control points available for selection at any given time. While the participant was generally able to maintain spatial awareness of control points within individual lenses, some confusion arose when interacting with multiple quadrics placed in close proximity or at small scales. 
\textbf{F$_{5}$} \textit{Standard DVR favored for spatial awareness}.
The DVR with depth culling received the lowest ratings. A major concern was the lack of occlusion cues needed to perceive the relative position of the quadric surface within the volume. It was also noted that neighboring depth culling could be confused with the inherent transparency of the volume rendering. Overall, the participant stated that the standard DVR provided the most effective contextual and depth cues while still allowing clear perception of the quadric surfaces in focus.

\vspace{3px}
\noindent \textbf{Future work} Given the expert’s preference for compound quadric surfaces, we are interested to investigate mechanisms for composing and editing multiple quadric images as a unified structure. They noted that aligning sequential quadrics to form a continuous surface could become time-consuming and result in visible gaps. As they expressed, it would be desirable ``\textit{to be able to connect geometries together}''. Similarly, they repeatedly adjusted curvatures at the joins between adjacent quadrics. Rather than manually bending each quadric one at a time, we plan on offering a smoother, continuous interaction model---free from intermittent breaks---to enhance fluidity and improve overall user performance.
Lastly, we have been encouraged to investigate ways to leverage the use of guides. This is envisioned for future design iterations---\textit{e.g.} users could be enabled to ``snap'' quadrics to descriptive lines within the guides, or even construct swept quadric surfaces directly from these reference lines, similar to the approach proposed by Jackson \textit{et al.}~\cite{Jackson:2016}.

\vspace{-4px}
\section{Conclusion}
\label{sec:conclusion}
\vspace{-3px}

We introduced a virtual reality lens technique for bringing heterogeneous features into focus in DVR visualizations.
We model the lens as a parametrizable quadric surface, intended to conform to a feature patch based on the premise that a faithful feature approximation can be built through the composition of multiple lenses.
We detail our visualization and interaction design, and discuss findings of an initial qualitative evaluation that not only showcases lens usage but also highlights opportunities for future research on lens visualization.

\bibliographystyle{abbrv-doi}

\bibliography{template}

\end{document}